\documentclass[10pt]{article}

\usepackage[margin=2cm]{geometry}
\usepackage{amsfonts}
\usepackage{graphicx} 
\usepackage{url}
\usepackage{subfig}
\usepackage{subcaption}
\usepackage{caption}
\usepackage{amsmath}
\usepackage{amssymb}
\usepackage{tensor}
\usepackage{color}
\usepackage{soul}
\usepackage[normalem]{ulem}
\usepackage{authblk}  
\usepackage{epsfig}
\usepackage{float} 
\usepackage[toc,page]{appendix}
\usepackage{pgfplots}
\usepackage{enumitem}

\usepackage{hyperref}
\usepackage[all]{hypcap}
\usepackage{cleveref}

 \usepackage[style=phys,biblabel=brackets,eprint=true,backend=bibtex]{biblatex}
\definecolor{DarkGreen}{RGB}{0,215,0}

\title{\bf There and back again: \\ Outspiraling motion in non-Kerr compact objects}

\author[1]{Manuel O. Mariano\footnote{mgomariano@ua.pt}}
\author[1,2]{C. A. R. Herdeiro\footnote{herdeiro@ua.pt}}

\affil[1]{\normalsize Departamento de Matemática da Universidade de Aveiro and
Center for Research and Development in Mathematics and Applications – CIDMA Campus de Santiago, 3810-183 Aveiro, Portugal}

\affil[2]{\normalsize Programa de Pós-Graduação em Física, Universidade Federal do Pará, 66075-110, Belém, Par{\'a}, Brazil }

\date{July 2025}

\begin{document}

\maketitle

\begin{abstract}
In Keplerian dynamics, a test body orbiting a point particle in circular motion has a monotonically increasing frequency, with decreasing radius. If a dissipative channel is introduced, such as gravitational wave (GW) emission, under the quadrupole approximation, the corresponding GW strain has an ever-increasing frequency with time. A similar statement holds for equatorial motion of a test particle on the Kerr manifold, except such inspiral is cut off at the ISCO, wherein stable circular orbits cease to exist and a plunge is expected. We analyze circular timelike orbits in generic spinning spacetimes and study the conditions in which \textit{exotic} motion can occur, arising from non-Kerr features. In particular, we derive conditions under which an inspiral towards a compact object is naturally followed by an \textit{outspiral} motion, and give concrete examples as well as the corresponding GW phenomenology. This analysis serves both as a theoretical exploration of non-Kerrness and as an example of a concrete smoking gun of exotic spacetimes.
\end{abstract}

\tableofcontents

\newpage

\section{Introduction}

The detection of $\mathcal{O}(100)$ Gravitational Wave (GW) events by the LIGO/Virgo/KAGRA collaboration \cite{LIGOScientific:2016aoc,LIGOScientific:2018mvr,LIGOScientific:2020ibl,KAGRA:2021vkt} from the coalescence of stellar mass black holes (BHs) and neutron stars has allowed for unprecedented tests of general relativity (GR) in the strong field regime \cite{LIGOScientific:2021sio}. In the future, space-based detectors such as the Laser Interferometer Space Antenna (LISA) \cite{LISA:2024hlh} will be observing GWs from supermassive BH mergers and highly asymmetric binaries, which will allow for the probing of GR against novel astrophysical phenomena \cite{LISA:2022yao,LISACosmologyWorkingGroup:2022jok, LISA:2022kgy, Caprini:2025mfr}.

Anchored on well-established uniqueness theorems \cite{Carter:1971ac,Robinson:1975bv,Chrusciel:2012jk}, GR suggests that all isolated astrophysical BH candidates, ranging (at least) 10 orders of magnitude in mass, are described by the Kerr solution. This is the \emph{Kerr Hypothesis}. Stable (on cosmological timescales) non-Kerr models with a plausible dynamical formation mechanism, however, can emerge naturally from minimally coupled GR to exotic matter or modified gravity theories \cite{Herdeiro:2022yle}, and challenge this hypothesis. Thus, it is crucial to explore and model smoking gun non-Kerr features of these objects, to then search for them in the detected GW signals \cite{Cardoso:2019rvt}.

The study of spacetime geodesics can play an important role in identifying non-Kerr signatures. In a metric theory of gravity, the orbits of a stellar or intermediate-mass compact object (the secondary) around a compact object with a much larger mass (the primary) are adequately described by geodesics of the primary spacetime. The geodesic structure for both Kerr \cite{Wilkins:1972rs,Chandrasekhar:1985kt} and a variety of non-Kerr spacetimes \cite{Diemer:2013zms,Grandclement:2014msa,Grould:2017rzz,Glampedakis:2018blj,Bacchini:2021fig,Collodel:2021gxu,Delgado:2021jxd,Destounis:2021mqv,Zhang:2021xhp,Zhang:2022qzw,Pombo:2023ody,Rosa:2024bqv,Sengo:2024pwk,Torres:2024eli,Gjorgjieski:2025uik,Bermudez-Cardenas:2025hrp} has been extensively studied in the literature. Examples of non-Kerr \textit{qualitative} geodesic features are: the occurrence of pointy-petal orbits and semi-orbits \cite{Grandclement:2014msa,Grould:2017rzz,Collodel:2017end}, static rings (SRs) and light points \cite{Collodel:2017end,Grandclement:2016eng}, disconnected regions of timelike stable circular orbits \cite{Sengo:2024pwk,Delgado:2021jxd} or even chaotic signatures due to nonintegrable geodesics \cite{Apostolatos:2009vu,Destounis:2021mqv}. These non-Kerr attributes may leave detectable phenomenological imprints \cite{Kesden:2004qx,Delgado:2023wnj,Sengo:2024pwk,Collodel:2021gxu,Collodel:2021jwi,Teodoro:2020kok,Teodoro:2021ezj,Destounis:2023khj,Rosa:2024bqv,Cipriani:2025ini}, for instance in the emitted gravitational waveform. Radiated GWs will carry energy away from the binary, making the secondary move in a ``quasi-geodesic" motion. Under the \emph{adiabatic approximation} \cite{Glampedakis:2002cb,Hughes:2001jr,Gair:2005is,Glampedakis:2002ya} a two-timescale approach is employed, whereby the secondary moves along geodesics with a constant orbital energy at timescales of the orbital period, only inspiraling inwards for much larger times. Although this approximation neglects backreacting self-force terms \cite{Poisson:2011nh}, it is commonly 
assumed in the study of extreme mass ratio inspirals (EMRIs), one of LISA's key targets \cite{Amaro-Seoane:2007osp,Babak:2017tow,Berry:2019wgg,Cardenas-Avendano:2024mqp}.

Particular examples of a non-Kerr GW signature was found in \cite{Collodel:2021gxu} and \cite{Delgado:2023wnj}, where equatorial quasi-circular EMRIs around Kerr BHs with scalar hair were studied. These are composite spinning horizon-boson star systems, that emerge as solutions of the Einstein-(complex)-Klein-Gordon theory \cite{Herdeiro:2014goa,Herdeiro:2015gia,Herdeiro:2015waa}. They can form dynamically through the superradiant instability \cite{Brito:2015oca,East:2017ovw,Herdeiro:2017phl}, provided that (in $G=c=1$ units) the Compton wavelength $1/\mu$ of the scalar field, with $\mu$ the boson's mass, is larger than the horizon scale set by the BH mass $\mathcal{M}$, i.e., $\mathcal{M}\mu\leq 1$. For some very hairy solutions, it was found that the GW signal can ``backward chirp", due to the nonmonotonicity of the metric functions for such spacetimes \cite{Herdeiro:2014jaa}.

For a Kerr BH primary, as the secondary radiates GWs in an equatorial quasi-circular motion, it will transition to circular geodesics with lower radii, until the innermost stable circular orbit (ISCO) is reached, wherein the secondary plunges. In \cite{Lehebel:2022yyz}, it was claimed 
that adiabatic loss of energy always causes the secondary to slowly inspiral towards the primary, ending either in a plunge (just like in Kerr) or on a SR that coincides with a minima of the generalized Newtonian potential. In this work, we claim that an extra possibility should also be considered: that it is possible for the secondary to reach an orbit where prograde and retrograde motions become degenerate, and energy loss leads to a continuous \emph{transition from an inspiraling to an outspiraling regime}. Moreover, this non-Kerr feature can be found in rotating spacetimes with nonmonotonic behavior of the metric functions. This can be realized, for instance, in off-centered mass distributions. Two concrete families of examples are discussed below.

This work is organized as follows. In Sec.\ref{section_generic_spacetimes}, we discuss equatorial circular geodesics for a generic $\mathbb{Z}_2$, asymptotically flat, circular and rotating spacetime. In Sec.\ref{section 3}, we consider an energy loss mechanism, via the emission of GWs under the quadrupole approximation, and set the conditions under which an outspiral motion is possible, determining also its possible outcomes. In Sec.\ref{sec4} we describe spacetimes where the outspiral is possible, and in Sec.\ref{sec5} we discuss the GW signatures of this regime on such spacetimes.

\section{Circular orbits in generic rotating spacetimes}
\label{section_generic_spacetimes}
Consider a compact object, with or without an event horizon, which is stationary, axisymmetric and asymptotically flat, in 1+3 dimensions. The two Killing vector fields $\chi^{\mu}$ and $\eta^{\mu}$ are associated, respectively, with stationarity and axisymmetry. Then it holds that $\left[\chi,\eta\right]=0$ \cite{Carter:1970ea}. Therefore, a coordinate system $(t,r,\theta,\varphi)$ adapted simultaneously to both symmetries  exists; $\chi=\partial_t$ and $\eta=\partial_{\varphi}$. 
Moreover, the spacetime is assumed circular: $\chi^{\mu} \tensor{R}{_{\mu}^{[\nu}}\chi^{\rho}\eta^{\sigma]}=\eta^{\mu} \tensor{R}{_{\mu}^{[\nu}}\chi^{\rho}\eta^{\sigma]}=0$, where $R_{\mu \nu}$ is the Ricci tensor. Physically, this implies that energy currents exist only along $\partial_\varphi$, with no convective or radial components \cite{Lehebel:2022yyz}.
For asymptotically flat spacetimes, circularity implies that the 2-planes orthogonal to $\eta^{\mu}$ and $\chi^{\mu}$ are integrable, i.e span a 2-surface \cite{Wald:1984rg}. Therefore, one can set the coordinates $(r,\theta)$ to parameterize the orthogonal 2-surface, as well as orthogonal amongst themselves \cite{Chandrasekhar:1985kt}. This yields the metric~\cite{Delgado:2021jxd}
\begin{equation}
    ds^2 = g_{tt}(r,\theta)dt^2+2g_{t\varphi}(r,\theta)dtd\varphi+g_{\varphi\varphi}(r,\theta)d\varphi^2+g_{rr}(r,\theta)dr^2+g_{\theta \theta}(r,\theta)d\theta^2.
\end{equation}

Observe that circularity implies the discrete symmetry $(t,\varphi) \rightarrow (-t,-\varphi)$.  Gauge freedom allows to set the event horizon, if it exists, at a constant (positive) radial coordinate $r=r_H$. Thus, the exterior spacetime is $r_H<r<\infty$ (or $0\leq r < \infty$ if no horizon exists). Asymptotically (at $r \rightarrow \infty$) our coordinates must match the standard spherical coordinates. Thus, the standard coordinate ranges $\theta \in [0, \pi]$, $\varphi \in [0,2\pi]$ and $t \in \mathbb{R}$ apply.
Finally, a $\mathbb{Z}_2$ symmetry, $\theta \rightarrow \pi - \theta$ is imposed, defining an ``equator" at $\theta = \pi/2$ as its set of fixed points. This is a totally geodesic submanifold. On this equator, the perimetral radius $R=\sqrt{g_{\varphi \varphi}}|_{\theta = \pi/2}$ is used,\footnote{Outside the horizon, causality requires that $g_{\varphi\varphi} \geq 0$.} such that the integral lines of $\partial_\varphi$ are circumferences with perimeter $2 \pi R$. Therefore, the equatorial plane metric becomes
\begin{equation}\label{metric}
    ds^2_{\theta=\pi/2} = g_{tt}(R,\pi/2)dt^2+2g_{t\varphi}(R,\pi/2)dtd\varphi+R^2d\varphi^2+g_{RR}(R,\pi/2)dR^2.
\end{equation}

Consider now circular equatorial timelike geodesics in this spacetime. The equations of motion can be obtained from the following Lagrangian $\mathcal{L}$:
\begin{equation}\label{lagrangian}
    \frac{2\mathcal{L}}{m} = g_{\mu \nu}\dot{x}^{\mu}\dot{x}^{\nu} = g_{tt}\dot{t}^2+2g_{t \varphi}\dot{t}\dot{\varphi}+ R^2\dot{\varphi}^2+g_{RR}\dot{R}^2=-1,
\end{equation}
where the dot denotes a derivative with respect to the proper time $\tau$, and $m$ is the mass of the secondary. To each Killing vector is associated a constant of motion
\begin{equation}\label{const1}
    E=-g_{\mu\nu}\dot{x}^{\mu}\chi^{\nu}=-g_{tt}\dot{t}- g_{t \varphi}\dot{\varphi},
\end{equation}
\begin{equation}\label{const2}
    L=g_{\mu\nu}\dot{x}^{\mu}\eta^{\nu}=g_{t\varphi}\dot{t}+ R^2\dot{\varphi}.
\end{equation}
These quantities are, respectively, the energy and angular momentum per unit mass of the secondary. Plugging \cref{const1,const2} into the Lagrangian in \cref{lagrangian}, we get
\begin{equation}\label{lagrangian2}
    g_{RR}\dot{R}^2=-1 + \frac{R^2 E^2 + 2 g_{t \varphi}EL+g_{tt}L^2}{B(R)} \equiv V(R,E,L),
\end{equation}
where
\begin{equation}
    B(R)\equiv g_{t\varphi}^2-g_{tt}R^2.
\end{equation}
The problem has been reduced to a particle moving in a 1D effective potential $V$, with $g_{RR}$ acting as a position-dependent mass term. In the domain of outer communications $B(R)>0$ due to the signature of the metric in the $(t,\varphi)$ sector.

To obtain $E\equiv E(R_{cir})$ and $L\equiv L(R_{cir})$ for each equatorial circular geodesic of radius $R_{cir}$, one needs to solve for two conditions~\cite{VanAelst:2020aws}. From \cref{lagrangian2} we obtain
\begin{equation}\label{circ1}
    V(R_{\text{cir}},E,L)=0. 
\end{equation}
Using the standard Euler-Lagrange equation for $R$ from the Lagrangian in \cref{lagrangian}, leads to
\begin{equation}\label{circ2}
   \Big(\frac{dV}{dR}\Big)_{R=R_{\rm cir}}\equiv V'(R_{\text{cir}},E,L)=0.
\end{equation}
To further impose that the circular orbit at $R=R_{\rm cir}$ is stable, one must ensure, additionally,
\begin{equation}\label{stable}
     \Big(\frac{d^2V}{dR^2}\Big)_{R=R_{\rm cir}}\equiv V''(R_{\text{cir}},E,L)<0.
\end{equation}

Together with \cref{lagrangian2}, that defines $V$, \cref{circ1,circ2} yield $E$ and $L$ for circular orbits. Due to the quadratic nature of the equations, one obtains for each $R_{cir}$ a pair of solutions corresponding to prograde (labeled $+$) and retrograde (labeled $-$) orbits. To write the solutions, it is useful to introduce the angular velocity of the secondary,
\begin{equation}
    \Omega \equiv \frac{d\varphi}{dt}=\frac{\dot{\varphi}}{\dot{t}}=-\frac{Eg_{t\varphi}-Lg_{tt}}{ER^2+Lg_{t\varphi}},
\end{equation}
where from now on we drop the subscript ``${\rm cir}$" in $R_{\text{cir}}$, but it is understood this refers to the radius of a timelike, circular equatorial orbit. The solution can be written as
\begin{equation}\label{sol1}
    E_{\pm}=-\frac{g_{tt}+g_{t\varphi}\Omega_{\pm}}{\sqrt{\beta_{\pm}}},
\end{equation}
\begin{equation}\label{sol2}
    L_{\pm}=\frac{g_{t\varphi}+R^2\Omega_{\pm}}{\sqrt{\beta_{\pm}}},
\end{equation}
\begin{equation}\label{sol3}
    \Omega_{\pm}=\frac{-g'_{t\varphi}\pm \sqrt{C}}{2R},
\end{equation} 
where we have defined
\begin{equation}\label{sol4}
    C \equiv (g'_{t\varphi})^2-2g'_{tt}R,
\end{equation}
and
\begin{equation} \label{sol5}
    \beta_{\pm}\equiv -g_{tt}-2g_{t\varphi}\Omega_{\pm}-R^2\Omega_{\pm}^2.
\end{equation}

As we will later see, retrograde (prograde) \textit{does not mean} corotating (counter-rotating) relative to the primary. The inequalities $\Omega_+>0$, $\Omega_-<0$, $L_+>0$ and $L_-<0$ are only true sufficiently far away from the primary. For timelike circular orbits, we must have $C\geq0$ and $\beta_{\pm}>0$. Below, we will mention what occurs when these inequalities break down. For now, observe that, at the radius where $C=0$, $\Omega_{+}=\Omega_{-}$ and so there is a degeneracy between prograde and retrograde orbits, implying only one possible circular geodesic motion. Finally, one can also express the second derivative of the potential as
\begin{equation}\label{potential}
    V''_{\pm}=\frac{\Omega_{\pm}^2 \gamma_{\varphi \varphi}+2\Omega_{\pm}\gamma_{t\varphi}+\gamma_{tt}}{B \beta_{\pm}},
\end{equation}
with
\begin{equation}
    \gamma_{\mu \nu}=g''_{\mu\nu}B+2g_{\mu\nu}C.
\end{equation}

\subsection{The case of Kerr}

To provide some intuition, we will now specify some of the above quantities in the case of Kerr. In Boyer-Lindquist (BL) coordinates $(t,r,\theta,\varphi)$, the Kerr metric takes the form 
\begin{equation}
    ds^2 = -\frac{\Delta}{\Sigma}\left(dt-a \sin^2\theta d\varphi\right)^2 + \frac{\Sigma}{\Delta}dr^2+\Sigma d\theta^2+\frac{\sin^2\theta}{\Sigma} \left[a dt - (r^2 +a^2)d\varphi \right]^2,
\end{equation}
with $\Sigma \equiv r^2+a^2\cos^2\theta$ and $\Delta\equiv r^2-2Mr+a^2$. The metric is described by two macroscopic parameters, the mass $M$ and the angular momentum $J=aM$. To avoid a naked singularity, the Kerr bound must be obeyed: $-1\leq j\leq 1$, with $j\equiv J/M^2$. It is more instructive, however, to write the metric in coordinates where $r \rightarrow \tilde r \equiv r/M$. Focusing on the equatorial plane alone ($\theta=\pi/2$), one obtains that

\begin{equation} \label{Kerr1}
    \widetilde{ g_{tt}} \equiv g_{tt}= 1-\frac{2}{\tilde r},
\end{equation}
\begin{equation}\label{Kerr2}
    \widetilde{g_{t\varphi}} \equiv g_{t\varphi} /M= -\frac{2j}{\tilde r},
\end{equation}
\begin{equation}\label{Kerr3}
    \widetilde{g_{\varphi\varphi}} \equiv g_{\varphi\varphi} /M^2 \equiv \tilde R^2 = \tilde r^2+j^2\left(1+\frac{2}{\tilde r}\right),
\end{equation}
with $\tilde R \equiv R/M$ the dimensionless perimetral radius and where the metric components with a tilde are also dimensionless. The relevant functions for equatorial, circular geodesics on the Kerr spacetime can be determined by using \cref{Kerr1,Kerr2,Kerr3} on \cref{sol1,sol2,sol3,sol4,sol5,potential}, yielding
\begin{equation}
    C = \frac{4}{\tilde r},
\end{equation}
\begin{equation}\label{KerrOmega}
    M\Omega_{\pm}= \frac{1}{j\pm \tilde r^{3/2}},
\end{equation}
\begin{equation}
    \beta_{\pm} =\frac{ (\tilde r-3)\tilde r^2 \pm 2j \tilde r^{3/2}}{(j\pm\tilde r^{3/2})^2},
\end{equation}
\begin{equation}
    E_\pm = \frac{(\tilde r-2) \tilde r\pm j\sqrt{\tilde r}}{\tilde r \sqrt{(\tilde r-3)\tilde r \pm 2j \sqrt{\tilde r}}},
 \end{equation}
\begin{equation}
    L_{\pm} = \pm \frac{\tilde r^2 \mp 2j\sqrt{\tilde r}+j^2}{\sqrt{\tilde r\left[(\tilde r-3)\tilde r \pm 2j\sqrt{\tilde r}\right]}},
\end{equation}
\begin{equation}
    V''_\pm = \frac{2\left[3j^2 \mp8j\sqrt{\tilde r}+(6-\tilde r)\tilde r\right]}{\left[\pm2j\tilde r^{3/2}+(\tilde r-3)\tilde r^2\right]\left[(\tilde r-2)\tilde r+j^2\right]}.
\end{equation}
The results are written in terms of the dimensionless radial BL coordinate $\tilde r$. To convert $\tilde r$ to $\tilde R$, one must solve the cubic equation from \cref{Kerr3}. Using Viete's solution \cite{Nickalls_2006}, and choosing the solution that matches $\tilde r = \tilde R$ when $j=0$, one finds that
\begin{equation}
    \tilde r = 2 \sqrt{\frac{\tilde R ^2 - j^2}{3}} \cos{\left(\frac{1}{3}\arccos{\left[ \sqrt{\frac{27}{(\tilde R ^2 - j^2)^3}}j^2  \right]}\right)}.
\end{equation}
In the limit $R\gg1$, the two radii also coincide. Moreover, at the outer horizon, where $\tilde r_H = 1 + \sqrt{1-j^2}$, one obtains that $\tilde R_H=2$ independently of the value of $j$.

From the above equations it follows that $C$ and $\Omega_\pm$ have no zeros; on the other hand, $\beta_\pm$ and $ V''_\pm$ have zeros. In the limit $j=0$, these are, respectively, at $\tilde{r}=3$ and $\tilde{r}=6$, which one identifies as the light ring (LR) and ISCO in Schwarzschild. As we shall see below, the zeros of these quantities are significant for the geodesic motion for generic spacetimes under the conditions specified in Sec.\ref{section_generic_spacetimes}.

\section{Evolution between circular orbits} \label{section 3}

\subsection{Adiabatic energy emission - Quadrupole waves}

Let us consider an energy dissipation mechanism allowing the secondary to evolve between timelike equatorial circular orbits. When the radiated energy during one orbit is much smaller than the orbital energy of the secondary, and thus the orbital period will be much shorter than the timescale of the inspiral, $T_{\text{inspiral}} \gg  T_{\text{orbit}}$, an adiabatic approximation holds. Then, the secondary motion can be approximated as a sequence of equatorial geodesics, with the energy and angular momentum fluxes (per unit mass), $dE/dt$ and $dL/dt$ respectively, dictating the transitions. Moreover, if the secondary follows a circular geodesic it will progress to another circular geodesic, following a quasi-circular motion, provided the fluxes obey
\begin{equation} \label{ang.mom}
    \frac{dL}{dt} = \frac{1}{\Omega}\frac{dE}{dt}.
\end{equation}
The backreaction from radiative mechanisms such as gravitational or electromagnetic radiation obeys \cref{ang.mom}, while environmental effects such as accretion or dynamical friction may, in general, lead to an increase of the orbital eccentricity \cite{Cardoso:2020iji}. Here, we shall consider the former only.

For quasi-circular motion, dissipated energy $\delta E <0$ leads to an adiabatic shift in the orbital radius $\delta R$: 
\begin{equation}\label{radius_energy}
    \delta E = E'\delta R \Rightarrow \frac{dE}{dt}=E'\frac{dR}{dt}.
\end{equation}
By differentiating \cref{sol1}, one arrives at \cite{Delgado:2023wnj,Lehebel:2022yyz}
\begin{equation}\label{master1}
    E'_{\pm}=\mp \frac{B\Omega_{\pm}V''_{\pm}}{2\sqrt{\beta_{\pm}C}}.
\end{equation}

Consider GW emission, modeled by the quadrupole approximation, as the energy dissipation mechanism.  Let the dissipated energy be normalized to the mass of the secondary, to match the orbital definition. Then, the radiated energy (per unit mass) flux is 
\begin{equation}\label{energy_loss1}
    \frac{d E}{dt}= -\frac{1}{5m} \left<{\frac{d^3 J_{ij}}{dt^3}\frac{d^3 J_{ij}}{dt^3}}\right>,
\end{equation}
with $J_{ij} = I_{ij} - \frac{1}{3} \delta^{ij} I_{ij}$, the reduced quadrupole moment and
\begin{equation}\label{quad.mom1}
    I_{ij} = \int y_i y_j \rho(t,\vec{y})  \,d^3y,
\end{equation}
where $\rho(t,\vec{y})$ is the energy density. Treating both the secondary and the primary as point-like masses, with the primary fixed at the origin and $\vec{y}_{\text{sec}}=(R \cos \Omega t, R \sin \Omega t, 0)$, yields
\begin{equation} \label{quad.mom2}
    \rho (t,\vec{y})=m \,\delta^{(3)}(\vec{y}-\vec{y}_{\text{sec}}).
\end{equation}
The primary contribution for $\rho$ was not included, as its position is fixed at the origin. The energy (per unit mass)  flux thus becomes
\begin{equation}\label{energy_loss}
    \frac{d{E}}{dt}=  -\frac{32}{5}m R^4\Omega^6,
\end{equation}
with the angular momentum flux obeying \cref{ang.mom}.

From \cref{master1,energy_loss}, the secondary's radial coordinate evolution due to GWs emission reads \cite{Delgado:2023wnj}
\begin{equation}\label{master}
    \frac{dR}{dt}=\pm\frac{64}{5}m R^4 \Omega_{\pm}^5\frac{\sqrt{\beta_{\pm}C}}{B V''_{\pm}}.
\end{equation}
Alternatively, using \cref{potential}, this can also be expressed as
\begin{equation}\label{master2}
    \frac{dR}{dt}=\pm\frac{64}{5}m R^4 \Omega_{\pm}^5\frac{\beta_{\pm}^{3/2}C^{1/2}}{\Omega_{\pm}^2 \gamma_{\varphi \varphi}+2\Omega_{\pm}\gamma_{t\varphi}+\gamma_{tt}}.
\end{equation}

\subsection{Critical points of $dR/dt$} \label{sec32}

Inspection of~\eqref{master} and~\eqref{master2} reveals that the behavior of $dR/dt$ has critical points at zeros of the functions $\Omega_\pm$, $V''_\pm$, $\beta_\pm$ and $C$. To contextualize, for equatorial circular orbits~\cite{Delgado:2021jxd}, the following meanings apply:
\begin{align} 
    V''<0 \ \ \ \ \ \ \ \ \ \ &\text{(Stability condition)}
    \label{condition1}\\
    \beta_{\pm}>0 \ \ \ \ \ \ \ \ \ \ &\text{(Timelike condition)} \label{condition2} \\
    C>0 \ \ \ \ \ \ \ \ \ \ &\text{(Existence condition)} \label{condition3} \ .
\end{align}

In the far region (from the primary), Keplerian mechanics,  and thus all these three conditions, must hold. Then, it is expected that as energy is lost via emission of gravitational waves, the object will transition to orbits with progressively lower radii. This is confirmed, e.g., from eq.~\eqref{master}. Indeed, prograde (retrograde) orbits obey $\Omega_{+}>0$ ($\Omega_{-}<0$). Thus, considering~\eqref{condition1}, in the far away region one gets  $ dR/dt<0$ in both cases.

In the near region, however, such behavior may end. The secondary may reach a last stable circular orbit continuously connected to spatial infinity. This is dubbed \emph{marginally stable circular orbit} (MSCO). For Kerr BHs, it coincides with the ISCO. In more generic compact objects, however, this need not be the case, and the specific behavior depends on the type of critical point for $dR/dt$, which we split into four distinct groups \cite{Delgado:2021jxd,Delgado:2023wnj}.

\begin{figure}[H]
    \centering
    \includegraphics[width=0.4\textwidth]{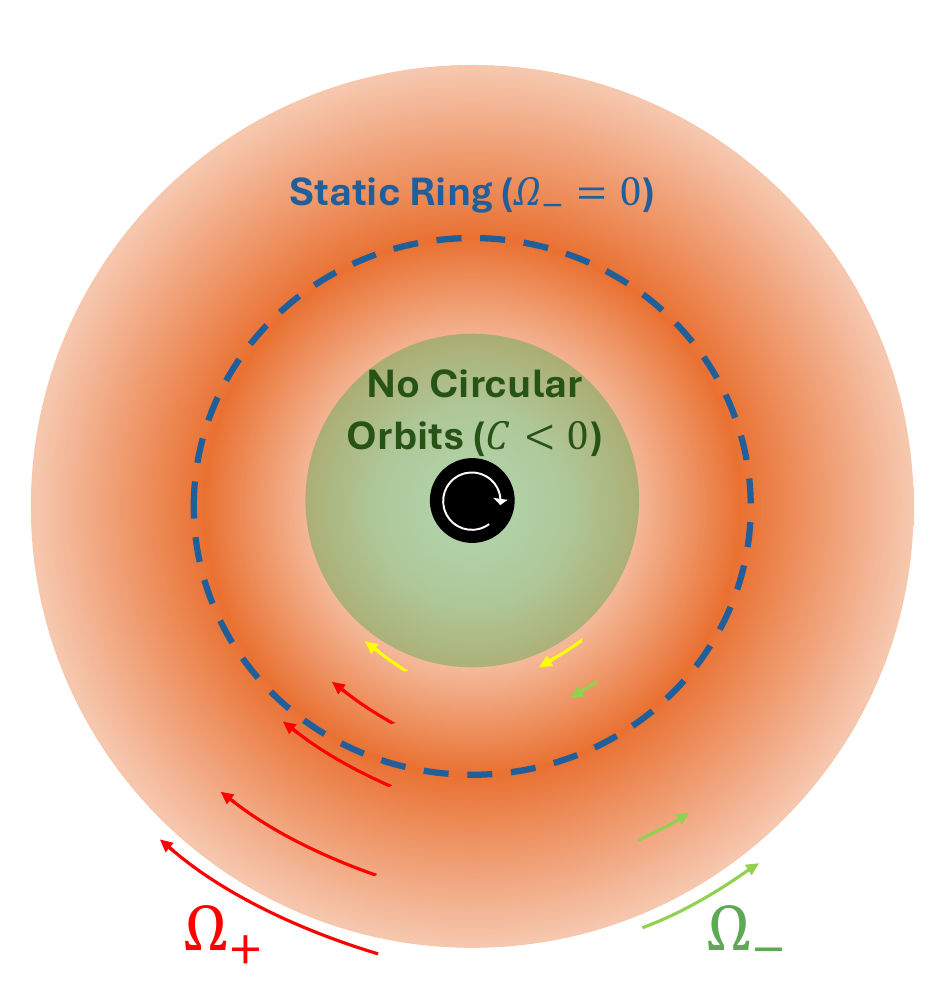}
    \hfill
    \includegraphics[width=0.55\textwidth]{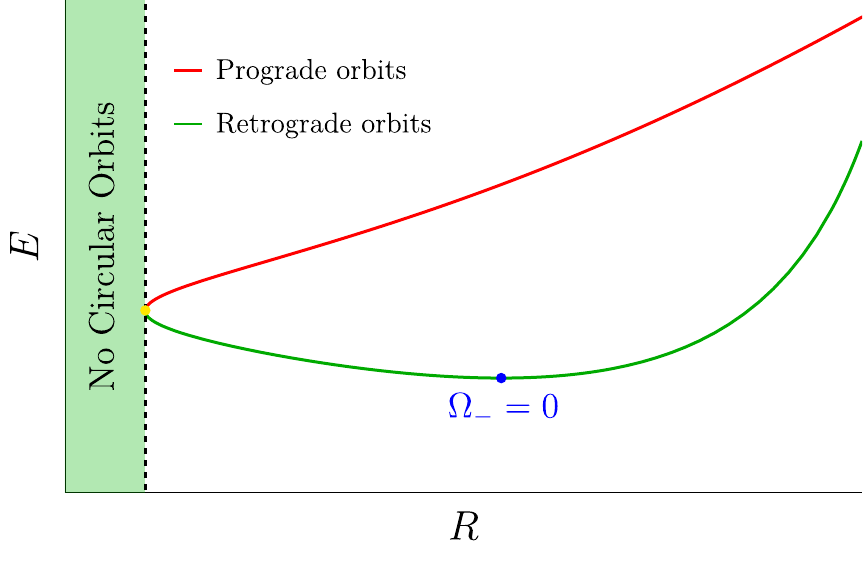}

    \caption{Illustration of the secondary motion with inspiral and outspiral. (Left) The spinning primary is envisioned as a core (black) and a surrounding environment (orange). Darker tones represent regions with higher density. A central (green) region has no circular orbits and a static ring (blue dashed line) is assumed for retrograde orbits. The arrows in red (green) represent qualitatively the evolution of the angular velocity for prograde (retrograde) orbits  $\Omega_+$ ($\Omega_-$), becoming degenerate at $C=0$ (yellow arrows), where $\Omega_+=\Omega_-$. (Right) Radial dependence of the circular orbits' energy. The yellow dot represents the radius at which $C=0$ and $E_+=E_-$. The secondary initially inspirals along prograde orbits, and then continues to lose energy by transitioning to a retrograde orbit at $R_C$ outspiraling towards the SR.}
    \label{fig:Scheme_outspiral}
\end{figure}

\begin{itemize}
    \item \textbf{$V_\pm''=0$}: \cref{master} predicts that $dR/dt\xrightarrow{}-\infty$ and thus a breakdown of the adiabaticity condition. Physically, this corresponds to the end of the \textit{stable} circular orbits region. Then, the secondary will enter a \emph{transition} regime. This may connect the inspiral and the plunge towards the primary, as in Kerr. But in other compact objects, one could envisage different possibilities as other, more interior, disconnected regions with stable timelike circular orbits could exist~\cite{Delgado:2021jxd,Sengo:2024pwk}.

    \item \textbf{$\beta_{\pm}=0$}: naively, the MSCO could be located at the outermost LR, determined by $\beta_{\pm}(R_{\text{LR}})=0$ \cite{Delgado:2021jxd}. Then, \cref{master} would predict $dR/dt\xrightarrow{}0$. For a vast class of ultracompact objects, however, with or without an event horizon, the outermost LR will be unstable \cite{Cunha:2017qtt,Cunha:2020azh,Delgado:2023wnj}. Timelike circular geodesics in the vicinity of an unstable LR, are either unstable or nonexistent \cite{Delgado:2021jxd}. Therefore, a larger radius where $V_\pm''=0$ should exist, leading to the previous case. 

    \item \textbf{$\Omega_{\pm}=0$}: this corresponds to a SR; with respect to an asymptotic static observer, an object following such an orbit would appear static. At the SR, the orbital energy simplifies to
    \begin{equation}
       (E_{\pm})_{\text{SR}}=\sqrt{(\beta_{\pm})_{\text{SR}}}=\sqrt{-g_{tt}},
    \end{equation}
    which corresponds to the local redshift factor. Thus a SR cannot occur inside an ergoregion. From \cref{sol3,sol4} $ \left(g'_{tt}\right)_{\text{SR}}=0$; thus SRs occur at local extrema of $g_{tt}$. Moreover, a SR is stable if $\left(g''_{tt}\right)_{\text{SR}}<0$.\footnote{For spherically symmetric spacetimes, this condition is sufficient \textit{and necessary} to ensure stability. In the rotating case, it is only sufficient.} This could naturally occur in objects with an off-centered matter distribution, e.g., spinning boson stars (BSs)~\cite{Collodel:2017end,Teodoro:2020kok}, Proca stars~\cite{Sengo:2024pwk}, or Kerr BHs with scalar hair~\cite{Delgado:2023wnj,Collodel:2021jwi}. Additionally, a SR coincides with a LR when $g_{tt}=g'_{tt}=0$. Such orbits are dubbed \emph{light points}, as photons would be at rest relative to an asymptotic observer \cite{Grandclement:2016eng}.

    As the secondary approaches a SR, \cref{master} predicts $dR/dt \rightarrow 0$. Thus, the SR represents the endpoint of the secondary inspiral evolution, acting as a \emph{floating orbit} of this spacetime, due to the energy flux $dE/dt\rightarrow 0$. From \cref{master1}, the SR coincides with an extreme of the orbital energy. If it is a minimum, it acts as an accumulation point for infalling matter. Even if the secondary would be initially placed at a
    circular orbit below the SR, it would actually outspiral towards the SR \cite{Delgado:2023wnj,Lehebel:2022yyz}.

    \item \textbf{$C=0$:} if $C(R_C)=0$ and $C(R)>0$ for $R>R_C$, then as $R\rightarrow R_C$, $dR/dt \rightarrow 0$,  from \cref{master}. Unlike the SR case, however, in this case $dE/dt = 0$ is not guaranteed by~\eqref{energy_loss}, suggesting the evolution must continue rather than attaining a floating orbit. Indeed an interesting turn of affairs ensues.  
     
     From \cref{sol1,sol2,sol3,sol4,sol5}, when $C=0$ both prograde and retrograde orbits have equal angular velocities and thus become degenerate. Therefore, at $R=R_C$, the secondary can actually \textit{transition} between the two types of orbits. Assume for concreteness $\Omega_+>0$ at $R=R_C$; a similar analysis holds for the other case. Then, from \cref{master}, in the prograde case, $dR/dt(R_C + \epsilon)<0$, ($\epsilon>0$) whereas in the retrograde case, $dR/dt(R_C + \epsilon)>0$. Thus a secondary arriving at $R_C$ in a prograde motion transitions to retrograde motion and \emph{outspirals}. Indeed,  from \cref{master1}, at $R=R_C + \epsilon$,  $E'_+>0$ and $E'_-<0$, such that energy can continue to be dissipated, as long as the secondary transitions from an inspiraling prograde orbit to an outspiraling retrograde orbit,  at $R_C$ - see Fig.~\ref{fig:Scheme_outspiral}.

\end{itemize}

\subsection{Duration of the inspiral and outcome of the outspiral}

The outspiral behavior just described raises two questions. Firstly, does the inspiral until $R_C$ occur in a finite time? Secondly, assuming the outspiral occurs, what is the endpoint of the evolution?

Let us address the first question. One can calculate the time $\Delta t$ taken to inspiral from an initial circular orbit at radius $R_0>R_C$, along prograde orbits,\footnote{\textit{Mutatis mutandis} for retrograde orbits.} by evaluating
\begin{equation} \label{duration}
    \Delta t = -\int_{R_C}^{R_0}\left[ \frac{64}{5}m r^4 \Omega_+^5\frac{\beta_+^{3/2}C^{1/2}}{\Omega_+^2 \gamma_{\varphi \varphi}+2\Omega_+\gamma_{t\varphi}+\gamma_{tt}} \right]^{-1}  dR \equiv -\int_{R_C}^{R_0} \frac{dR}{f(R)},
\end{equation}
where the function $f(R)$ corresponds to the righthand side of \cref{master}. At $R_C$,  $f\rightarrow 0 $, and the convergence of \cref{duration} needs to be studied with care. Expanding $f$ around its zero $R_C$, using an ordinary Taylor series, leads to problems as $f'\rightarrow + \infty$ at $R_C$. Yet, one can expand $f^2$, thus yielding
\begin{equation}
    f(R) = \underbrace{\left(\frac{64}{5}m R^4 \Omega_+^5\frac{\beta_+^{3/2} }{\Omega_+^2 \gamma_{\varphi \varphi}+2\Omega_+\gamma_{t\varphi}+\gamma_{tt}} \left(\frac{dC}{dR}\right)^{1/2}\right)_{R=R_C}}_{\equiv\, C_0} \sqrt{R-R_C}  + \mathcal{O}(\sqrt[3]{R-R_C}).
\end{equation}
For $R<R_C$ ($R>R_C$), $C<0$ ($C>0$); then $dC/dR>0$ at $R_C$. From this expansion, one can then decompose the integral into a sum of two terms
\begin{equation} \label{duration2}
    \Delta t = -\left(\int_{R_C+\epsilon}^{R_0}\frac{dR}{f(R)} \, + \frac{1}{C_0}\int_0^{\epsilon}\frac{d   \epsilon'}{\sqrt{\epsilon'}} \right),
\end{equation}
with $\epsilon$ some arbitrarily small positive quantity, allowing us to disregard all higher-order terms in the second term. From \cref{duration2}, it is then clear that both pieces converge, and so \emph{the secondary will reach $R_C$ in a finite amount of time}, proceeding then to outspiral along retrograde orbits.

Consider now the second question: the possible endpoints of the outspiral regime.  As $\Omega_+=\Omega_-$ at $R=R_C$, while near spatial infinity $\Omega_-<0$ and $\Omega_+>0$, there will always be a zero of either $\Omega_+$ or $\Omega_-$ between $R_C$ and infinity. For concreteness, assume $\Omega_-$ is the one vanishing. Similar considerations apply to the other case. Then, there are several possibilities:
\begin{enumerate}
    \item [i)] This zero of $\Omega_-$ is a SR \textit{and} it is continuously connected to $R_C$ by stable timelike retrograde orbits. That is, conditions \eqref{condition1} and \eqref{condition2} both hold between $R_C$ and $R_{\rm SR}$. Then, the outspiral will end at the SR. Computing $\Delta t$ for such outspiral between $R_C$ and $R_{\rm SR}$, considering that $f \sim (R-R_{\rm SR})^5$ in the SR's vicinity \cite{Delgado:2023wnj}, one observes that \emph{the secondary will reach the SR in an infinite amount of time}.\footnote{This actually applies regardless of whether the approach is an inspiral or an outspiral motion.}  Globally, if \eqref{condition1} and \eqref{condition2} hold between $R_C$ and spatial infinity, SRs are the accumulation point for infalling matter starting arbitrarily far from the primary and \emph{regardless of its initial rotation sense} (see Fig.~\ref{fig:Scheme_outspiral}), albeit reaching the SR takes an infinite time.

\item [ii)] This zero of $\Omega_-$  is not continuously connected to $R_C$ by stable timelike retrograde orbits because either \eqref{condition1} or \eqref{condition2} is violated in between. In the latter case, a stable LR exists at $R_{LR}$ and stable timelike circular orbits exist for $R_C<R<R_{LR}$ \cite{Delgado:2021jxd}. Thus, the secondary will tend towards the LR, albeit taking an infinite time. In the former case, the secondary will outspiral towards a radius wherein $V''=0$, delimiting a region of unstable circular orbits. Such an \emph{outspiral plunge} signals a breakdown of the adiabaticity condition.
\end{enumerate}

\section{The quasi-circular outspiraling - concrete illustrations}
\label{sec4}
Let us consider concrete primary spacetimes where the discussed outspiraling occurs. In subsection \ref{sec41}, a simple toy model is considered, while in subsection \ref{sec42}, our focus turns to spinning BSs. In Sec.\ref{sec5}, \cref{master} is solved for both these families of examples to look for the GW signatures of such motion.

\subsection{Engineered geometry} \label{sec41}
We construct a simplified \textit{ad hoc} geometry, within the symmetries discussed in Sec.\ref{section_generic_spacetimes}, with the desired geodesic behavior. To determine the relevant functions [see \cref{sol1,sol2,sol3,sol4,sol5,potential}], it is enough to define a $g_{tt}$ and $g_{t\varphi}$ exclusively on the equator. The following conditions are imposed:

\begin{enumerate}[label=\roman*.]
    \item $g_{t\varphi}=0=g'_{t\varphi}=g'_{tt}$ at $R=0$;
    
    \item $g_{tt}\rightarrow-1+{\alpha}/{R}$ and  $g_{t\varphi}\rightarrow-{\beta}/{R}$ at $R\rightarrow \infty$, with $\alpha,\beta>0$; 
    
    \item there exists $R_C>0$ such that $C(R_C)=0$, and $C(R)>0$ for all $R>R_C$;

    \item $\beta_{\pm}>0$ and $V''_{\pm}<0$, for all $R\geq R_C$.
\end{enumerate}
Requirement i) ensures regularity at the origin, while ii) enforces the right asymptotic behavior. Conditions iii) and iv) guarantee that the desired outspiral orbits for the secondary are possible. Observe that iii) is verified if $g_{tt}$ has a local maximum between $R_C$ and infinity.

\begin{figure}[H]
    \centering
    \includegraphics[width=0.43\textwidth]{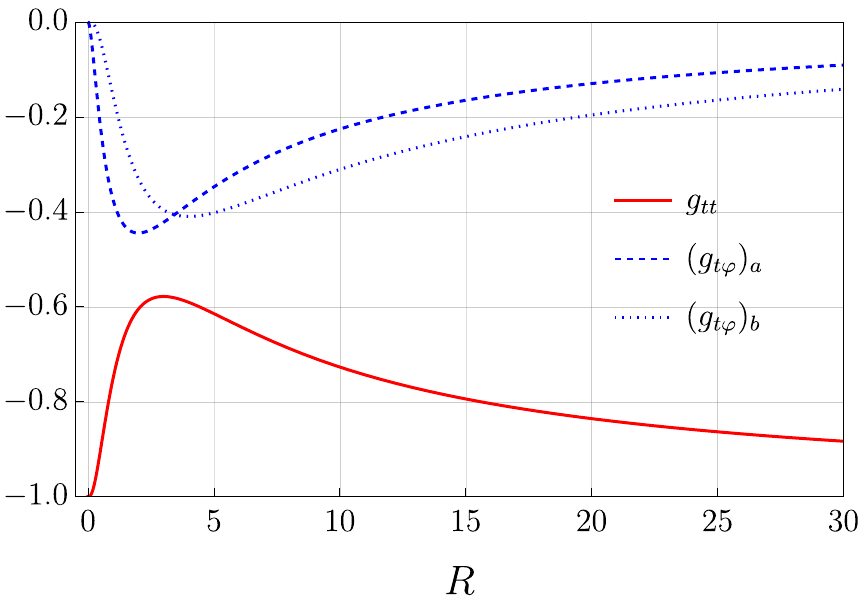}
    \hfill
    \includegraphics[width=0.48\textwidth]{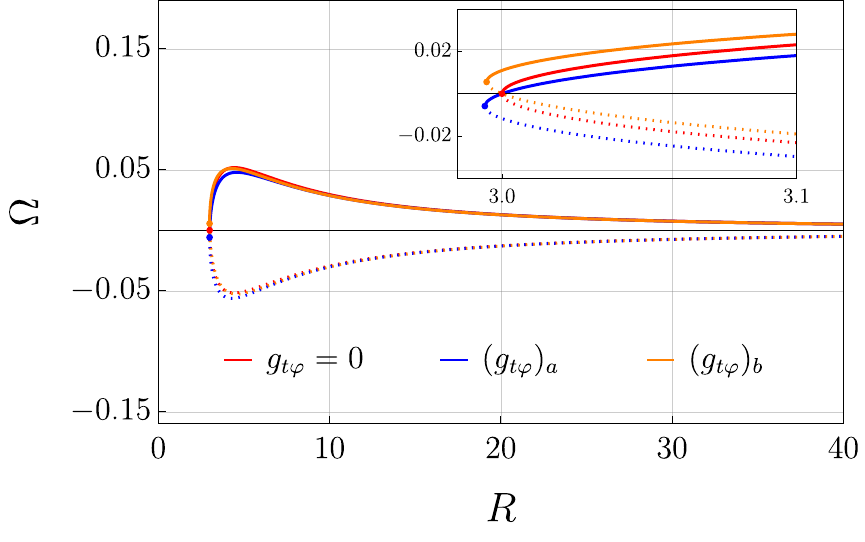}

    \caption{Toy model results. (Left) Radial profiles of $g_{tt}$ and $g_{t\varphi}$ in \cref{toy_metric} are parameterized by $a_1=4$, $k_1=3$ for $g_{tt}$; $a_2=2$, $k_2=4$ for $(g_{t\varphi})_a$; and $a_2=2$, $k_2=4$ for $(g_{t\varphi})_b$. (Right) Radial profile of $\Omega$ for the three distinct $g_{t\varphi}$ profiles. The continuous (dotted) lines correspond to prograde (retrograde) orbits, while the dots mark $C=0$. The inset plot shows the curves zoomed near $R=R_C$ }
    \label{fig:toymetric}
\end{figure}

The following parameterized choices are able to meet the requirements under appropriate choices of parameters:
\begin{equation}\label{toy_metric}
    g_{tt}=-1+\frac{a_1R^{k_1}}{(1+R)^{k_1+1}}\; \ , \qquad \;
    g_{t\varphi}=-\frac{a_2R^{k_2}}{(1+R)^{k_2+1}} \ ;
\end{equation}
for $ g_{tt}$ we choose $a_1=4$ and $k_1=3$; for $g_{t\varphi}$ three sets of constants are chosen: a) $a_2=3$ and $k_2=2$; b) $a_2=2$ and $k_2=4$; c) $a_2=0$. With the two former choices, both $g_{tt}$ and $g_{t\varphi}$  have one local extremum, with case a)/b) setting the local minimum of $g_{t\varphi}$ at a larger/lower radius than the local maximum of $g_{tt}$ - see left panel of Fig.~\ref{fig:toymetric}. Case c) is a spherically symmetric geometry, with $g_{t\varphi}=0$.

In the spherically symmetric setting, $R_C$ is a SR, $cf.$~\cref{sol3}. This degeneracy is lifted once $g_{t\varphi} \neq 0$, allowing an outspiraling region. According to Fig.~\ref{fig:toymetric} (right panel), in case a) the SR is retrograde with the outspiraling regime achieved if the secondary approaches the primary via prograde orbits. In case b) the SR is prograde and thus the outspiral occurs for initially retrograde orbits. This is a consequence of the relative position of the extrema of $g_{tt}$ and $g_{t\varphi}$ or, in other words, of the sign of $g'_{t\varphi}$ at $R_{\rm SR}$. 

At first glance, SRs could be interpreted as a balance between the secondary's angular momentum  and the primary's frame dragging \cite{Collodel:2017end,Grandclement:2016eng}. If true, SRs would always be retrograde, as in case a). Cases b) and c), however, contradict such interpretation~\cite{Lehebel:2022yyz}. Nonetheless, for the spinning BSs example only retrograde SRs are found.

\subsection{Spinning boson stars} \label{sec42}

BSs are horizonless, globally regular, asymptotically flat, equilibrium solutions of the Einstein-Klein-Gordon theory with a complex massive scalar field $\Psi$ minimally coupled to Einstein's gravity. We shall consider only mini-BSs, where no self-interactions exist. This model is described by the action 
\begin{equation} \label{action}
    S = \int \,d^4x \sqrt{-g}\left[ \frac{\mathcal{R}}{16 \pi} - g^{\alpha \beta} \partial_{\alpha}\Psi^{*} \partial_{\beta}\Psi - \mu^2 \Psi^{*} \Psi\right],
\end{equation}
with $\mathcal{R}$ the Ricci scalar, $\mu$ the scalar field mass and $\Psi^*$ the complex conjugate of $\Psi$. The metric ansatz considered to obtain the stationary and axisymmetric solutions reads
\begin{equation}\label{metric_ansatz}
    ds^2 = -e^{2 F_0(r,\theta)} dt^2 + e^{2F_1(r,\theta)}\left(dr^2 + r^2 d\theta^2\right) + e^{2F_2(r,\theta)}r^2 \sin^2\theta \left(d\varphi - \frac{W(r,\theta)}{r}dt\right)^2, 
\end{equation}
while the scalar field obeys
\begin{equation}\label{scalar_ansatz}
    \Psi(t,r,\theta,\varphi) = \phi(r,\theta) e^{i( m\varphi-\omega t)},
\end{equation}
with $\omega$ being the frequency of the scalar field, and $m=0,\pm1,\pm2, \ldots$ being the azimuthal harmonic index with $m=0$ corresponding to spherically symmetric solutions. Moreover, the solutions are also labeled by the non-negative integer $n$ associated with the number of radial nodes of $\phi$. The solutions considered here are characterized by $m=1$ and $n=0$ and were obtained numerically through the use of the CADSOL package \cite{SCHONAUER1988185,SCHONAUER2001473}. One important feature of these solutions is that boundary conditions, concerning the regularity of $\Psi$ at the origin, impose that $\phi |_{r=0}=0$ \cite{Kleihaus:2007vk}. This imposes a toroidal, off-centered distribution of the energy density. Observe that the scalar field mass $\mu$ can be absorbed into the coordinates, by performing the substitution $r\rightarrow r\mu$ and $\omega \rightarrow \omega/\mu$. Furthermore, $\omega/\mu$ can be interpreted as a label that identifies uniquely a mini-BS, as only first branch solutions are considered.

\begin{figure}[H]
    \centering
    \includegraphics[width=0.48\textwidth]{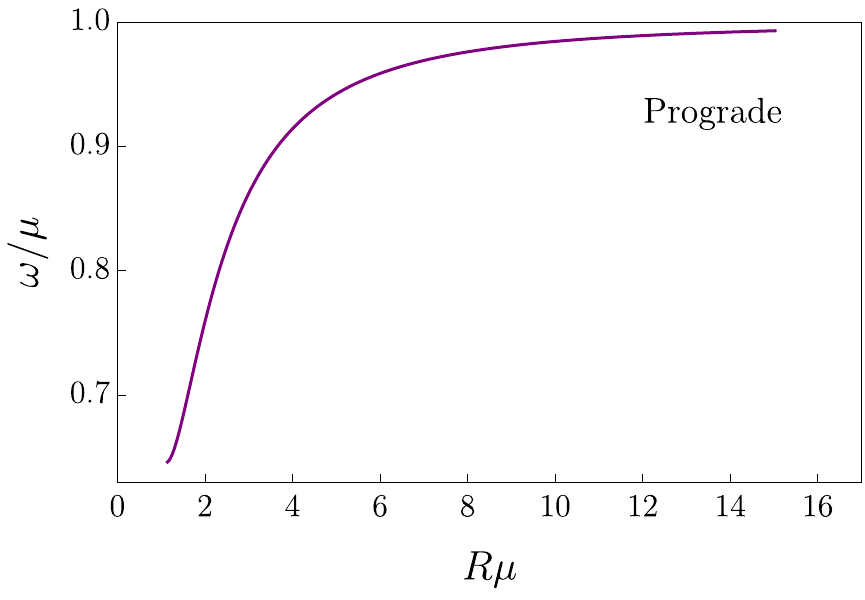}
    \hfill
    \includegraphics[width=0.48\textwidth]{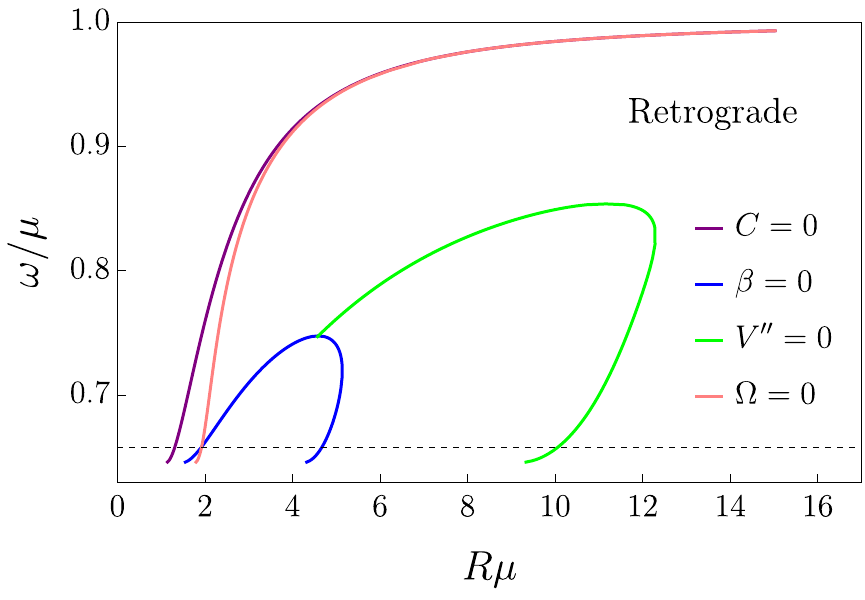}

    \caption{Structure of prograde (left) and retrograde (right) equatorial circular orbits for rotating BSs in the first branch. The black vertical dashed line on the right panel corresponds to $\omega/\mu=0.658$, from which the endpoint switches from a SR to a stable LR.}
    \label{fig:boson_star_stability}
\end{figure}

The results for the orbital stability regions for prograde and retrograde orbits can be found in Fig.~\ref{fig:boson_star_stability}. The stability of the two sets of orbits differ greatly, with only retrograde orbits displaying LRs and SRs. The radius of the $C=0$ orbit is common to both, as it is independent of the rotation sense. For these solutions, an outspiraling motion can exist exclusively for initially prograde orbits.

\begin{figure}[H]
    \centering
    \includegraphics[width=0.48\textwidth]{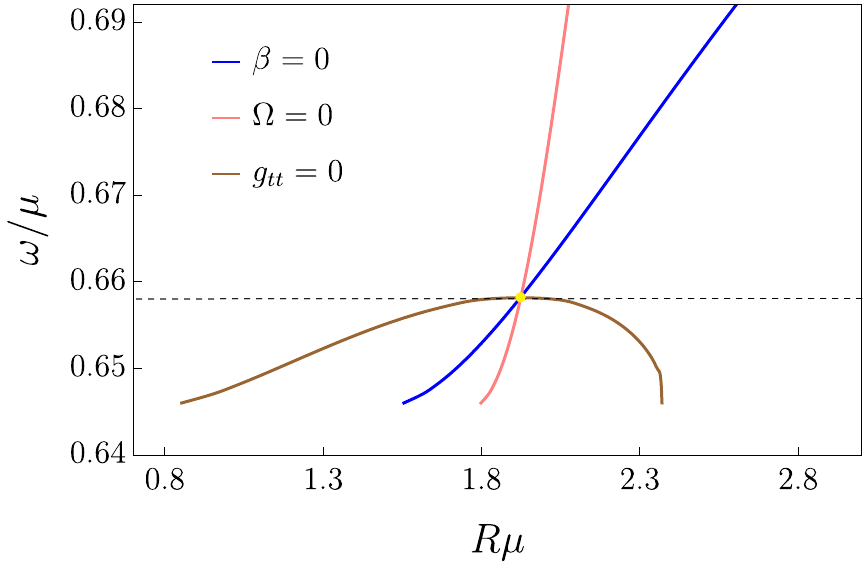}
    \hfill
    \includegraphics[width=0.48\textwidth]{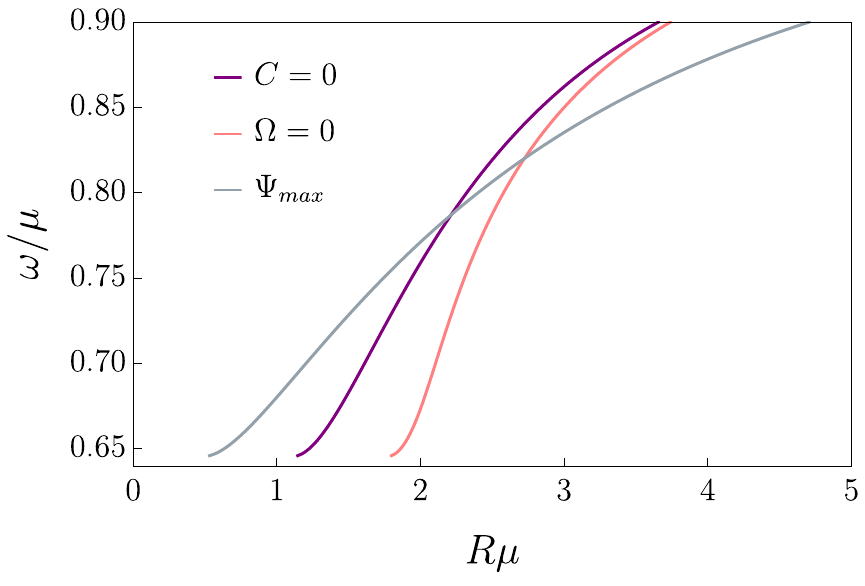}

    \caption{Properties of the rotating BSs. (Left) Emergence of ergoregion from light point orbit, where an ergosurface, a stable LR and the SR coincide. The horizontal dashed line corresponds to $\omega/\mu=0.658$, with the yellow dot representing the light point orbit. (Right) Plot of the radial coordinate where $|\Psi|$ is maximum with the SR and $C=0$.}
    \label{fig:bs_properties}
\end{figure}

For this class of examples, there are two outcomes for the outspiral motion. For $\omega/\mu>0.658$, the endpoint is the SR, while for $\omega/\mu<0.658$  the endpoint is the stable LR. For the solution with $\omega/\mu=0.658$ the SR coincides with the stable LR, leading to a light point orbit - Fig.~\ref{fig:bs_properties} (left panel). This spinning BS also marks the emergence of an ergoregion, present for all solutions with $\omega/\mu<0.658$~\cite{Grandclement:2016eng}. Therefore, all BSs where outspiraling towards a stable LR occurs are affected by the \emph{ergoregion instability}. Also,  the stable LR, although being retrograde, will be corotating with the primary for all BSs with an ergoregion. Finally, we observe there are both BSs where the maximum of the scalar field occurs at a larger and lower radius than both the SR and the $C=0$ orbits, dismissing an obvious correlation between these quantities - Fig.~\ref{fig:bs_properties} (right panel).

\section{Gravitational Wave Signatures} 
\label{sec5}
Having laid out the theoretical description of the possible motions in generic spacetimes, and presented concrete illustrations of the outspiraling motion, it is of interest to examine possible phenomenological implications. To achieve that, we shall now  solve numerically \cref{master} for the examples in Sec.\ref{sec4}. As a technical point on solving this equation, due to $dR/dt=0$ at $R=R_C$, to ``kick-start" the outspiral once this radius is attained from the inspiral, a perturbation is imposed taking the secondary to $R_C+\epsilon$. Then we obtain the GW strain as follows. In the quadrupole limit, the transverse traceless (TT) part of the metric perturbation (i.e. strain, assumed to be in the wave zone) at distance $D$ is
\begin{equation} \label{Quad1}
    h^{TT}_{ij} = \frac{2}{D} \frac{d^2 J^{TT}_{ij}}{dt^2},
\end{equation}
with $J_{ij}$ being the reduced quadrupole moment. Assuming the primary is head-on with the GW detector, it holds that $J^{TT}_{ij}=J_{ij}$. From \cref{quad.mom1,quad.mom2}, the two gauge-invariant degrees of freedom of $h^{TT}_{ij}$, the polarizations $h_{\text{+}} $ and $h_{\times}$, are 
\begin{equation}\label{pol1}
    h_{\text{+}} = -\frac{4 m}{D} R^2 \Omega^2 \cos 2\Omega t,
\end{equation}
\begin{equation}\label{pol2}
    h_{\times} = -\frac{4 m}{D} R^2 \Omega^2 \sin 2\Omega t.
\end{equation}
In the quadrupole limit, the GW frequency is simply twice the angular velocity, i.e $f_{GW}=2\Omega$, and
\begin{equation}\label{amplitude}
   \frac{h_{+}D}{m}=4R^2 \Omega^2.
\end{equation}
Therefore, the righthand side in \cref{amplitude} will be proportional to the strain amplitude.

\subsection{Engineered example}
We first display the results for the engineered metrics considered in subsection \ref{sec41}. The mass factor $m$ in \cref{master} was absorbed by the dimensionful quantities: $R, \Omega $ and $t$. The initial conditions considered were $R/m = 10$ at $t=0$.

\begin{figure}[H]
    \centering
    \includegraphics[width=0.48\textwidth]{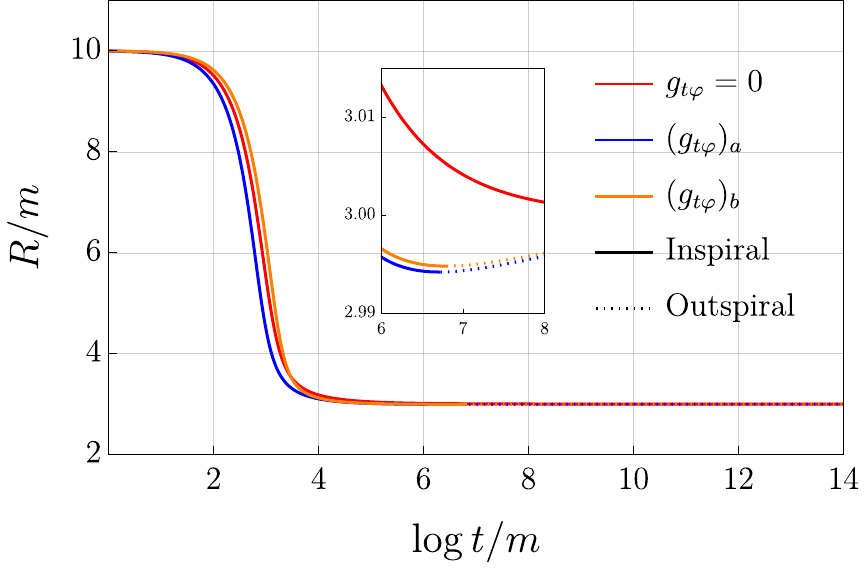}
    \hfill
    \includegraphics[width=0.48\textwidth]{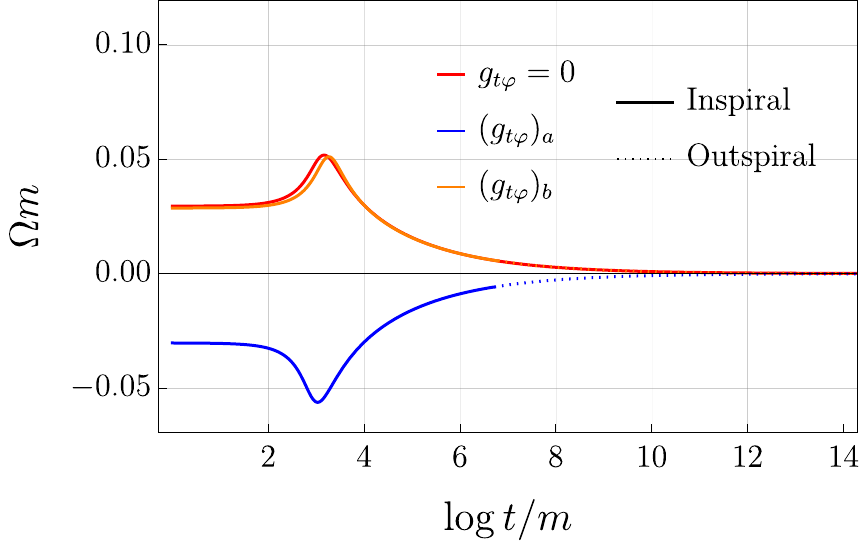}

    \caption{Evolution of $R /m$ and $\Omega m$ as a function of time for the engineered geometries. The parameters associated with the $g_{tt}$, $(g_{t\varphi})_a$, and $(g_{t\varphi})_b$ profiles of the underlying geometries can be found in the caption of Fig.~\ref{fig:toymetric}.}
    \label{fig:r_omega_evolution_toy}
\end{figure}

In the left panel of Fig.~\ref{fig:r_omega_evolution_toy} is displayed the outspiral behavior for spacetimes a) and b), occurring after the secondary reaches $R_C$, and ending towards the SR. By contrast, the spherically symmetric geometry displays only the inspiral directly towards the SR. In the right panel of Fig.~\ref{fig:r_omega_evolution_toy}, the angular frequency evolution is displayed. For all spacetimes, one observes a local maximum of $|\Omega|$ occurring during the inspiral phase. This would impose a backward chirp on the emitted GWs, similar to that found in \cite{Collodel:2021jwi,Delgado:2023wnj}. This qualitative feature is a consequence of the SR endpoint. Indeed, for $R\rightarrow + \infty$, one must have a monotonically increasing $|\Omega|$ as $R$ decreases, since $|\Omega| \sim 1/R^{3/2}$ [cf. \cref{KerrOmega}]. However, at the SR, $|\Omega|=0$. Indeed, this is a clear deviation from the Keplerian intuition, as for inspiraling orbits close to $R_C$ the angular frequency does not increase monotonically with decreasing radius. Thus $|\Omega|$ must attain a maximum at some radius and then decrease as it approaches the SR, giving rise to a backwards chirp. This is illustrated in the right panel of Fig.~\ref{fig:toymetric}. 
Consequently, the backwards chirp of the radiated GWs will always occur when a transition between an inspiral and outspiral is present, with the frequency maximum 
occurring before the transition point. By contrast, the existence of a backwards chirp does not imply an outspiral.

\begin{figure}[H]
	\centering
	\includegraphics[scale=0.6]{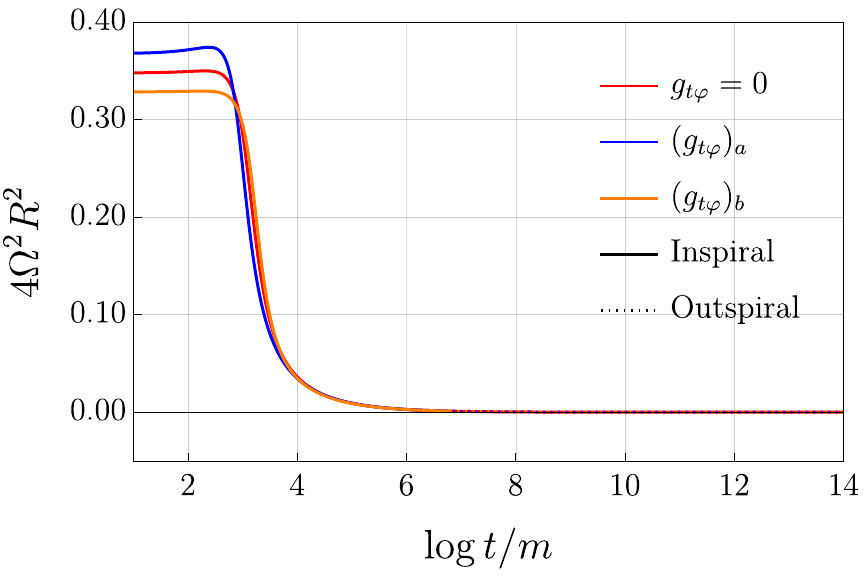}
     \caption{Evolution of $4R^2 \Omega^2 \propto h_{+}$ as a function of time for the engineered geometries. The parameters associated with the $g_{tt}$, $(g_{t\varphi})_a$, and $(g_{t\varphi})_b$ profiles of the underlying geometries can be found in the caption of Fig.~\ref{fig:toymetric}.}
     \label{fig:gw_evolution_toy}
\end{figure}

Finally, in Fig.~\ref{fig:gw_evolution_toy}, the evolution of the GW strain is displayed. As the secondary approaches the SR, $h_{+} \rightarrow0$ quadratically with $\Omega$. Thus, once the outspiral begins and the frequency is already decreasing towards $0$, the GW strain is already much smaller than the peak value achieved during the inspiral. These examples suggest that GW detectors may not be sensitive enough to detect radiation emitted during the outspiral, albeit a general proof is lacking.

\subsection{Spinning boson stars}

Let us now turn to some illustrative spinning BSs. The secondary is initially placed at a radius $R\mu=10$, and the mass ratio of the system is fixed at $m/M=10^{-5}$. Note that \cref{master} is invariant under the simultaneous transformation $t \rightarrow \alpha t$ and $m/M \rightarrow m/(\alpha M)$. So increasing or decreasing $m/M$ only changes the time duration of the inspiral in this approximation, leaving its dynamics intact. 

\begin{figure}[H]
    \centering
    \includegraphics[width=0.48\textwidth]{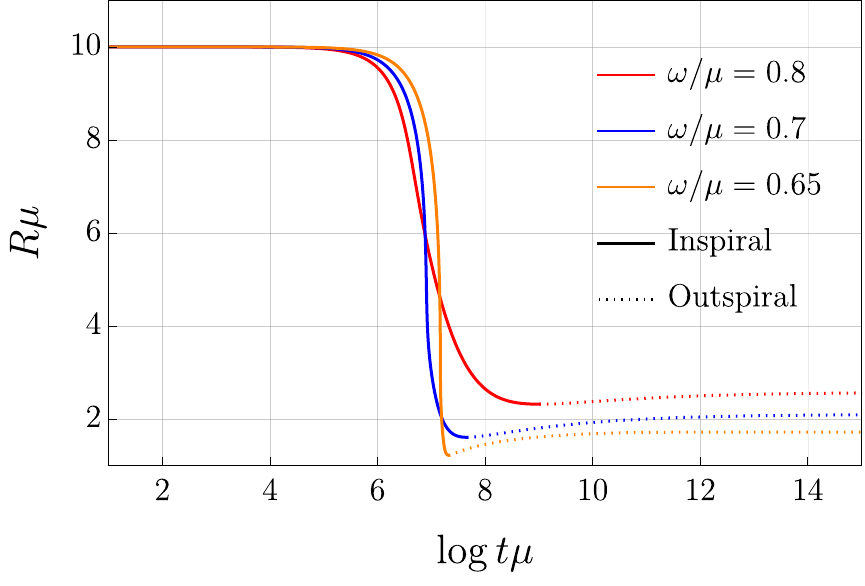}
    \hfill
    \includegraphics[width=0.48\textwidth]{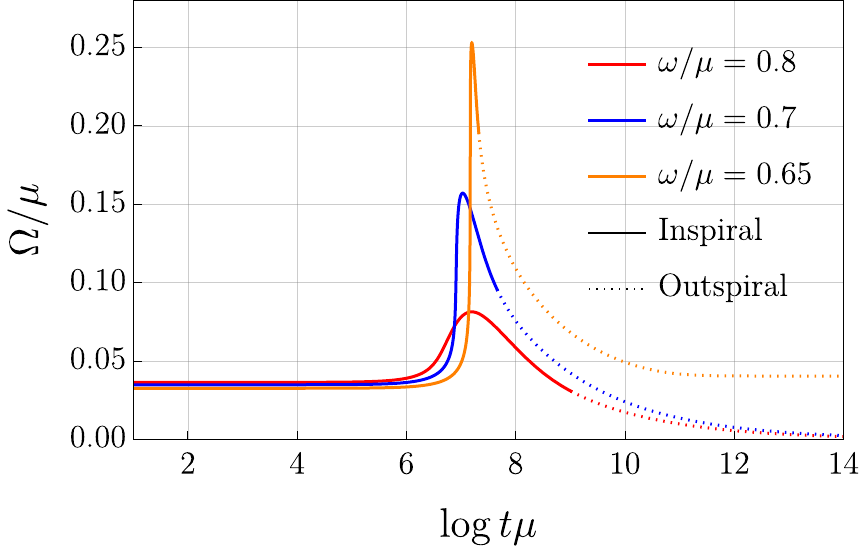}

    \caption{Evolution of $R \mu$ and $\Omega/\mu$ as a function of time for three distinct spinning BSs.}
    \label{fig:r_omega_evolution}
\end{figure}

The evolution was performed for rotating BSs in the first branch with $\omega / \mu=0.8,0.7,0.65$. In Fig.~\ref{fig:r_omega_evolution}, it is possible to see the time evolution of $R \mu$ and the orbital angular frequency $\Omega / \mu$ as a function of time. We see the outspiraling in the left panel and a backwards chirp in the frequency evolution in the right panel, just like in the prior subsection. The maximum of the angular frequency, once again, does not coincide with the beginning of the outspiral, occurring  slightly before, which is clear from the right panel.

\begin{figure}[H]
	\centering
	\includegraphics[scale=0.6]{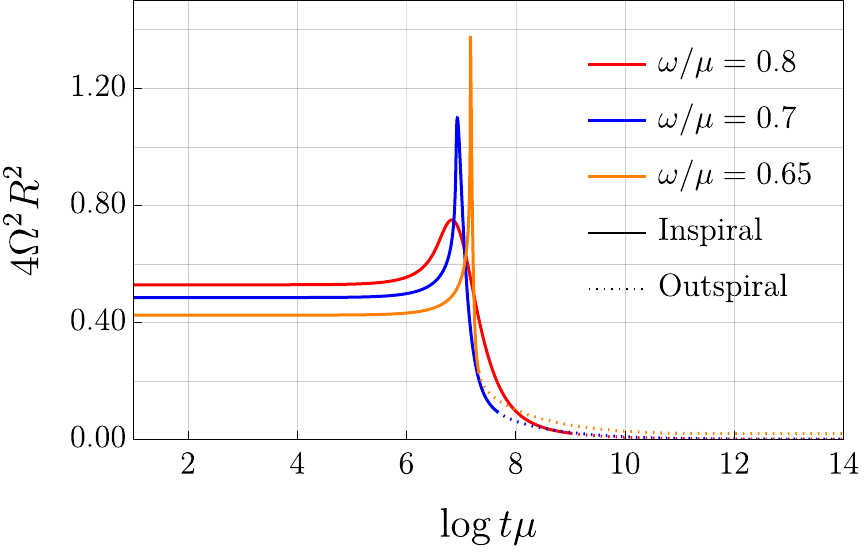}
     \caption{Evolution of $4R^2 \Omega^2 \propto h_{+}$ as a function of time for three distinct spinning BSs.}
     \label{fig:gw_evolution}
\end{figure}

We can now easily observe the behavior of the strain, from \cref{amplitude}. The righthand side of the latter equation is plotted in Fig.~\ref{fig:gw_evolution}, for the three considered BSs. Because $\Omega$ reaches a local maximum, the amplitude also peaks during the inspiral, and is considerably suppressed in the outspiral regime. For the $\omega / \mu=0.65$ BS, the outspiral endpoint is a stable LR, and so neither $\Omega /\mu$ nor $4R^2 \Omega^2$ vanish, even for large times.

\section{Conclusions and Discussions}

In this work, we determined the conditions under which an object orbiting in an equatorial quasi-circular motion, can continuously transition from an inspiral to an outspiral as it loses energy adiabatically. For this to occur, it is crucial that the spacetime is spinning and that $g_{tt}$ does not vary monotonically. By analyzing \cref{master}, derived assuming the quadrupole approximation for the radiated energy of GWs, we concluded that the transition can occur at a radius where a degeneracy between prograde and retrograde orbits exists, with the secondary transitioning between the two types of orbits. Moreover, we determined that such an orbit, albeit a stationary point of \cref{master}, could be reached in a finite amount of time. The outspiral endpoints could be either a SR, a stable LR, or a brief ``outwards plunge" as it enters a region of unstable timelike circular orbits. We displayed a couple of examples featuring this outspiral motion: an engineered geometry and a spinning BS. The outspiral endpoint for the engineered geometry was the SR, which could occur in either the retrograde or prograde branches. For spinning BSs, even though most cases featured a SR endpoint, we found some examples where the endpoint is a stable LR, albeit being also prone to an ergoregion instability. Additionally, no correlation between the maximum of the scalar field distribution with either the inspiral-outspiral transition radius or the SR location was found. Finally, we also calculated some GW signatures associated with the outspiral feature for both spacetimes. We concluded that a backwards chirp of the gravitational waveform was always present, even though the latter does not imply outspiraling. Indeed, the maximum of the secondary's angular velocity (or the GW frequency, under the quadrupole approximation) occurred before the transition. The GW strain during the outspiral motion was found to be much smaller, sometimes even negligible, relative to the maximum value.

There are a few additional remarks to be considered.  Even though \cref{master} is specialized for the GW emission under the quadrupole approximation, it is expected that the outspiral motion would still occur for other energy-dissipation formulae that tend to circularize orbits, i.e., that obey \cref{ang.mom}. The stationary behavior at $R_C$ does not come from the energy flux formula in \cref{energy_loss}, but from the circular geodesic structure of the spacetime in \cref{master1}. Thus, if another energy flux would be considered, such that $\dot{E}\neq 0$ at $R_C$, a transition would still be expected. We also remark that although the slow rotation approximation of BSs in~\cite{Delgado:2024fhc} reproduces nicely the total ADM mass and angular momenta, it cannot capture the outspiral behavior displayed in Fig.~\ref{fig:r_omega_evolution}. This is due to this approximation being unable to reproduce the $C<0$ region for the structure of circular orbits (c.f Fig.~\ref{fig:boson_star_stability}). Additionally, we note that the outspiraling described in this work represents a specific instance of a broader phenomenon that can arise for other physical setups. Examples include binaries undergoing mass loss \cite{Linial:2017hep,Kavic:2019cgk,Bambi:2025wjx}, systems with electromagnetic or scalar energy fluxes at play \cite{Santos:2024okf}, collisions of Proca stars in antiphase \cite{Palloni:2025mhn}, and scenarios in alternative theories of gravity where a small black hole is emitted by a larger one \cite{Islam:2024wqf}.

Many questions remain for future works. Firstly, it would be interesting to extend this formalism to eccentric orbits. In this case, the secondary would not ``see" an infinite potential barrier at $R_C$, but still the dissipation would be forcing the orbits to circularize [due to \cref{ang.mom}]. Additionally, we remark that the stable LR endpoint, albeit only reached at $t \rightarrow \infty$, appears to signal a breakdown of the formalism. At the stable LR, $\dot{E}\neq 0$, and thus as the secondary would approach the LR, an arbitrarily large amount of energy would be radiated via GWs, which is unphysical. Moreover, as the stable LR is being approached from inside, outwards of the stable LR, no timelike circular orbits are allowed, preventing further quasi-circular outspiral. Thus, the fate of such orbits approaching the stable LR from inwards is unclear. Finally, it would also be interesting to study the potential implications of the SR attractive nature as accumulation points.

\section{Acknowledgements}

This work is supported by CIDMA under the Portuguese Foundation for Science and Technology (FCT, \url{https://ror.org/00snfqn58}) Multi-Annual Financing Program for R\&D Units, grants UID/4106/2025 and UID/PRR/ 4106/2025. M.O.M. is supported by the FCT grant 2024.03055.BD.

\printbibliography[heading=bibintoc]

\end{document}